\documentclass[twocolumn,showpacs,preprintnumbers,aps, prb,amsmath,amssymb]{revtex4}
\usepackage[dvips]{graphicx}
\begin{document}
\title{
Stability of fixed points
in the $(4+\epsilon)$-dimensional random field O($N$) spin model 
for sufficiently large $N$
}

\author{Yoshinori Sakamoto}
\email{yossi@phys.ge.cst.nihon-u.ac.jp}
\affiliation{
Laboratory of Physics, College of Science and Technology, 
Nihon University, 7-24-1 Narashino-dai, Funabashi-city, Chiba, 
274-8501 Japan
}

\author{Hisamitsu Mukaida}
\email{mukaida@saitama-med.ac.jp}
\affiliation{
Department of Physics, Saitama Medical College, 
981 Kawakado Iruma-gun, Saitama, 350-0496 Japan
}

\author{Chigak Itoi}
\email{itoi@phys.cst.nihon-u.ac.jp}
\affiliation{
Department of Physics, College of Science and Technology, 
Nihon University, 1-8-14 Kanda-Surugadai, Chiyoda-ku, Tokyo, 
101-8308 Japan
}

\date{\today}

\begin{abstract}
We study the stability 
of fixed points in the two-loop renormalization group
for the random field O($N$) spin model in $4+\epsilon$ dimensions. 
We solve the fixed-point equation in the $1/N$ expansion 
and $\epsilon$ expansion. 
In the large-$N$ limit,
we study the stability of all fixed points.
We solve the eigenvalue equation 
for the infinitesimal deviation from the fixed points 
under physical conditions on the random anisotropy function.
We find that the fixed point corresponding to dimensional reduction 
is singly unstable and others are unstable or unphysical. 
Therefore, one has no choice other than 
dimensional reduction in the large-$N$ limit.
The two-loop $\beta$ function enables us to find a compact area 
in the $(d, N)$ plane where the dimensional reduction 
breaks down. 
We calculate higher-order corrections in the $1/N$ and $\epsilon$ 
expansions to the fixed point. 
Solving the corrected eigenvalue equation nonperturbatively, we find that 
this fixed point is singly unstable also for sufficiently large $N$
and the critical exponents 
show a dimensional reduction. 
\end{abstract}


\maketitle

\section{Introduction}
The random field O($N$) spin model 
is one of the simplest models with both quenched disorder 
and spin correlations. It is a fundamental problem 
to clarify the critical phenomena in this model. 
Dimensional reduction \cite{PS} is one key to solve this problem. 
Dimensional reduction claims that 
the critical behavior of the random field O($N$) spin model 
in $d$ space dimensions 
is the same as that of the pure O($N$) spin model in $d-2$ space dimensions. 
If this conjecture is true, all critical exponents 
of the random field spin model in $d$ dimensions 
should be identical to those of the corresponding pure model 
in $d-2$ dimensions.

Since several rigorous results for the random field Ising model 
($N=1$ case) indicated the failure of dimensional reduction to 
predict the lower critical dimensions, \cite{I,BK,A}  
the breakdown of dimensional reduction with some approximation methods 
was discussed in order to obtain intuitive understanding or
quantitative information. 
Fisher calculated the one-loop renormalization group, 
and he pointed out the breakdown of dimensional reduction 
due to the appearance of an infinite number of relevant operators 
in $4+\epsilon$ dimensions. \cite{Fi} 
He showed the existence of a fixed point corresponding to the dimensional reduction 
for $N\ge18$, but he argued that this fixed point should be unstable as far as the number 
of spin components, $N$, is finite. 
Therefore, he concluded that the dimensional reduction was not valid 
near four dimensions. Feldman found a 
nonanalytic fixed point in Fisher's 
renormalization group for several small $N$. \cite{Fe} 
He obtained nontrivial critical exponents shifted from 
the predictions of dimensional reduction.
M\'ezard and Young also suggested the breakdown of dimensional reduction 
by replica symmetry breaking. \cite{MY} 
Now, many researchers believe that dimensional reduction 
is incorrect in dimensions lower than 6. 

Recently, Tarjus and Tissier studied the critical phenomena of this model 
in any dimensions and for any value of $N$ 
by using the nonperturbative renormalization group method 
and the replica method. \cite{TT} 
They show the following relation of the critical exponents of the two-point 
spin correlation function: 
\begin{equation}
\eta=\bar{\eta}=\frac{\epsilon}{N-2},
\label{DR}
\end{equation}
predicted by dimensional reduction in a 
certain region in the $(d, N)$ plane. 
Since this relation seems valid for $N \geq 18$ near four dimensions, 
the consistency between their result and that of the Refs. \onlinecite{Fi} 
and \onlinecite{MY} should be studied. 

To understand the consistency of their works, the 
de Almeida--Thouless criterion \cite{AT} 
is applied faithfully to this model in a simple 
$1/N$-expansion method. \cite{SMI} It is shown that the 
saddle point of the auxiliary field is stable 
in the random field O($N$) spin model.
Also the stability argument by Balents and Fisher for random media \cite{BF} 
is applied to Fisher's one-loop renormalization group; then, 
the following two possibilities are indicated. \cite{SMI} 
The fixed point corresponding to dimensional reduction 
is singly unstable, or there is no singly unstable fixed point. 
Combining these results leads to relation (\ref{DR})
which is the same result obtained by Tarjus and Tissier.

In this paper, we study the stability of the fixed point
corresponding to dimensional reduction. Particularly,
we discuss the physical condition on the deviation from the fixed point. 
Since Fisher did not solve the eigenvalue problem completely
for the stability around the fixed point, \cite{Fi}
we solve this problem in a $1/N$ expansion. 
In the large-$N$ limit, the stability of all fixed points 
is studied. The solution of 
eigenvalue equations for the deviation from the fixed points 
indicates that the only once unstable fixed point 
shows the critical exponents predicted by dimensional reduction. 
Next, we study this fixed point 
by the two-loop renormalization group 
obtained by Le Doussal and Wiese \cite{DW}
and Tissier and Tarjus. \cite{TT2} 
The double expansion in $1/N$ and $\epsilon$
enables us to calculate the correction to the fixed point. 
The solution of the eigenvalue equation 
shows that the unstable modes pointed out by Fisher are fictitious. 
Therefore, we conclude that
the fixed point yielding relation (\ref{DR})
is singly unstable for sufficiently large $N$. 
This result agrees with that obtained by Tarjus and Tissier \cite{TT}
and also by the simple $1/N$ expansion. \cite{SMI} 
Furthermore, we calculate higher-order corrections to the eigenvalues.

This paper is organized as follows. In Sec. \ref{review}, 
we briefly review Fisher's renormalization group analysis 
for the random field O($N$) spin model in $4+\epsilon$ dimensions. \cite{Fi}  
In Sec. \ref{singularity}, we discuss possible 
singularities of the random anisotropy
functions
for a physical model.
In Sec. \ref{Large N limit}, we treat the one-loop renormalization group
in the large-$N$ limit. 
In this limit, 
the critical phenomena in $4+\epsilon$ dimensions are shown to 
be governed by the fixed point which 
gives the result of dimensional reduction. 
In Sec. \ref{Subleading corrections}, 
we investigate the stability of this fixed point 
corresponding to dimensional reduction 
by the two-loop renormalization group. 
We show that this fixed point is singly 
unstable on the basis of the physical condition on the 
coupling function discussed in Sec. \ref{singularity}. 
Thus, we conclude that the prediction of dimensional reduction
for the critical exponents (\ref{DR}) holds for sufficiently large $N$.
In Sec. \ref{higher}, we calculate higher-order corrections to the eigenvalue
and the exponents of the singularities.
In Sec. \ref{Conclusion}, we summarize our results and discuss some problems.

\section{Fisher's one-loop renormalization group analysis }
\label{review}

In this section, 
we briefly review Fisher's argument on the instability of 
the fixed point corresponding to the dimensional reduction
in the one-loop renormalization group
for the random field O($N$) spin model in $4+\epsilon$ dimensions. \cite{Fi}

\subsection{Model}
We consider O($N$) classical spins ${{\mbox{\boldmath$S$}}}(x)$ 
with a fixed-length constraint ${{\mbox{\boldmath$S$}}}(x)^2=1$. 
To take the average over the random field, 
one introduces replicas ${{\mbox{\boldmath$S$}}^{\alpha}}(x)$, 
$\alpha=1, \ldots, n$. 
We start from a nonlinear $\sigma$ model 
of the following replica partition function and action: 
\begin{eqnarray}
{\cal{Z}}
&=&
\int\prod_{\alpha=1}^n{\cal{D}}{\mbox{\boldmath$S$}}^{\alpha}
\delta({{\mbox{\boldmath$S$}}^{\alpha}}^2-1)
e^{-\beta H_{\rm rep}},\nonumber\\
\beta H_{\rm rep}
&=&
\frac{a^{2-d}}{2T}\int d^dx\sum_{\alpha =1}^n \sum_{\mu=1}^d
(\partial_{\mu}{\mbox{\boldmath$S$}}^{\alpha})^2\nonumber\\
&&-\frac{a^{-d}}{2T^2}\int d^dx\sum_{\alpha,\beta}^n 
R({\mbox{\boldmath$S$}}^{\alpha}\cdot{\mbox{\boldmath$S$}}^{\beta}),
\label{action}
\end{eqnarray}
where $a$ is the ultraviolet cutoff and 
the parameter $T$ denotes the dimensionless temperature. 
The function 
$R({\mbox{\boldmath$S$}}^{\alpha}\cdot{\mbox{\boldmath$S$}}^{\beta})$ 
represents general anisotropy 
including the random field and all the random anisotropies, 
and is given by 
\begin{eqnarray}
R({\mbox{\boldmath$S$}}^{\alpha}\cdot{\mbox{\boldmath$S$}}^{\beta})
=\sum_{\mu=1}^{\infty}\Delta_{\mu}
({\mbox{\boldmath$S$}}^{\alpha}\cdot{\mbox{\boldmath$S$}}^{\beta})^{\mu},
\end{eqnarray}
where $\Delta_{\mu}$ denotes 
the strength of the random field and the $\mu$th rank random anisotropy 
($\mu=1$ is the random field, 
and $\mu\ge2$ is the second- and higher-rank random anisotropy). 
These coupling constants are positive semidefinite $\Delta_\mu \geq 0$.

\subsection{One-loop $\beta$ function}
\begin{figure}[t]
\begin{center}
\setlength{\unitlength}{1mm}
\begin{picture}(80, 40)(0,0)
     \put(0,-5){ 
\includegraphics[width=80mm]{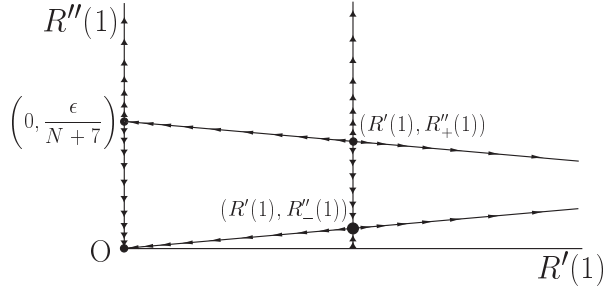}
		}
\end{picture}
\end{center}
\caption{The renormalization group flow for the couplings 
$R'(1)$ and $R''(1)$. The fixed point $(0, \frac{\epsilon}{N+7})$ is unphysical, because the 
corresponding $\Delta_1$ becomes negative. See Sec. \ref{Fixed points}.  }
\label{flow2}
\end{figure}

The $\beta$ function $\partial_tR(z)$ at zero temperature 
can be expressed in the loop expansion 
\begin{equation}
\partial_tR(z) = \beta_0[R] +\beta_1[R]+\beta_2[R] +\cdots.
\end{equation}
Here, we have defined
the scale parameter $t$ which increases toward the infrared direction.
Fisher calculated the one-loop $\beta$ function in the following form
\begin{eqnarray}
\beta_0[R]&=&-\epsilon R(z), \\
\beta_1[R]&=&2(N-2)R'(1)R(z)
-(N-1)zR'(1)R'(z)\nonumber\\
&&+(1-z^2)R'(1)R''(z)+\frac{1}{2}R'(z)^2(N-2+z^2)\nonumber\\
&&-R'(z)R''(z)z(1-z^2)+\frac{1}{2}R''(z)^2(1-z^2)^2.
\label{1loop}
\end{eqnarray}
Expanding $R(z)$ around $z=1$, we obtain the one-loop $\beta$ functions 
for $R'(1)$ and $R''(1)$: 
\begin{eqnarray}
\partial_tR'(1)
&=&
-\epsilon R'(1)+(N-2)R'(1)^2,\\
\label{beta1}
\partial_tR''(1)
&=&
-\epsilon R''(1)+6R'(1)R''(1)\nonumber\\
&&+(N+7)R''(1)^2+R'(1)^2.
\label{beta2}
\end{eqnarray}
The $\beta$ functions (\ref{beta1}) and (\ref{beta2}) 
have two nontrivial fixed points (see Fig. \ref{flow2}): 
\begin{eqnarray}
&&(R'(1),R_+''(1))\nonumber\\
&&=\biggl(
\frac{\epsilon}{N-2},
\frac{(N-8)+\sqrt{(N-2)(N-18)}}{2(N-2)(N+7)}\epsilon
\biggr),
\label{2+}\\
&&(R'(1),R_-''(1))\nonumber\\
&&=\biggl(
\frac{\epsilon}{N-2},
\frac{(N-8)-\sqrt{(N-2)(N-18)}}{2(N-2)(N+7)}\epsilon
\biggr).
\label{2-}
\end{eqnarray}
The formulas for the critical exponents $\eta$ and $\bar{\eta}$, 
\begin{eqnarray}
\eta=R'(1),\quad{\bar{\eta}}=(N-1)R'(1)-\epsilon, \label{RGcharge}
\end{eqnarray}
enable us to obtain the correlation function critical exponents (\ref{DR}).
This result confirms one of the predictions by dimensional reduction. 
From the fixed points (\ref{2+}) and (\ref{2-}), 
we find that these results are applicable only for $N\ge18$. 
The eigenvalues $\epsilon \lambda_1$ and $\epsilon \lambda_2^{\pm}$ of 
the scaling matrix at the fixed points (\ref{2+}) and (\ref{2-}) 
are given by
\begin{eqnarray}
\lambda_1&=&+1,\\
\lambda_2^{\pm}&=&\pm\sqrt{\frac{N-18}{N-2}}.
\end{eqnarray}
Thus, the fixed point (\ref{2+}) is unstable. 
The fixed point (\ref{2-}) seems to be stable for $N\ge18$. 
%

However, Fisher claimed an instability of 
this fixed point (\ref{2-}). 
For the one-loop $\beta$ function in terms of 
differential coefficients $R^{(k)}(1)$, $(k= 0, 1, 2, \ldots)$,
he obtained a triangular scaling matrix at the fixed point 
$(R'(1)^*, R_-''(1)^*, \ldots, R^{(k-1)}(1)^*, \ldots)$, whose 
diagonal components were given in the following series: 
\begin{eqnarray}
\lambda_k
&=&\frac{2k^2-k(N-1)+2N-4}{N-2}-1+ kNR_-''(1)^*\nonumber\\
&\simeq&
1-k+\frac{2k^2}{N} \ \ \ ( k \ge 3 ).
\label{eigen k}
\end{eqnarray}
Almost all eigenvalues were indicated to be positive 
for a sufficiently large $k$, 
although one should add a term $2 n k P_2 P_k$ missed in Eq. (C6) of his paper. \cite{Fi} 
Then, Fisher concluded that there was no singly unstable fixed point 
and dimensional reduction broke down near four dimensions. 
In Sec. \ref{Subleading corrections}, however,
we show that these infinitely many relevant modes
are unphysical by solving this eigenvalue problem completely. 
In the following sections, we carefully reexamine the renormalization group.

\section{Allowed singularities of the random anisotropy function}
\label{singularity}

The fixed-point condition of the renormalization group determines properties of the function $R(z)$.
Here we discuss possible asymptotic behaviors of $R(z)$ near $z=\pm 1$. 
The first derivative of the fixed-point equation with respect to $z$
is
\begin{eqnarray}
&&[-\epsilon +(N-3)R'(1)]R'(z) +z R'(z)^2\nonumber\\
&&-(N+1)zR'(1)R''(z)+(N-3+4z^2)R'(z)R''(z)\nonumber\\
&&+(1-z^2)R'(1)R'''(z)-z(1-z^2)R'(z)R'''(z)\nonumber\\
&&-3z(1-z^2)R''(z)^2+(1-z^2)^2R''(z)R'''(z)=0.
\label{fp'}
\end{eqnarray}
If we assume the asymptotic behavior of $R'(z)$ near $z=1$, 
\begin{equation}
R'(z) = R'(1)+ C(1-z)^\alpha +  \cdots,
\end{equation}
with $0 < \alpha$.  To discuss a cuspy behavior of $R(z)$ at $z=1$,
we consider only $\alpha < 1$.
The condition (\ref{fp'}) gives the following constraint: 
\begin{eqnarray}
&&[-\epsilon + (N-2)R'(1)]R'(1)\nonumber\\
&&-C^2 \alpha(4 \alpha^2+4\alpha+N-1)(1-z)^{2\alpha-1}=0.
\end{eqnarray}
For $\alpha \neq 1/2$, this constraint gives 
\begin{eqnarray}
\alpha = \frac{1}{2}(-1+\sqrt{2-N})
\end{eqnarray}
or 
\begin{eqnarray}
C=0, 
\label{nosing}
\end{eqnarray}
and also 
\begin{eqnarray}
&&R'(1)=\frac{\epsilon}{N-2}
\label{dr}
\end{eqnarray}
or 
\begin{eqnarray}
R'(1)=0.
\end{eqnarray}
Here, the former case given by Eqs.(\ref{nosing}) and (\ref{dr}) 
shows the dimensional reduction.
The formulas for (\ref{RGcharge}), 
the critical exponents obtained by Feldman, \cite{Fe}
enable us to obtain the critical exponents 
$$
\eta= \frac{\epsilon}{N-2}= \bar{\eta}.
$$
Therefore, no $\alpha \neq 1/2$ is allowed for any $N >2$. 
For $\alpha=1/2$, the parameter $R'(1)$ can change continuously 
depending on the constant $C$. 
Therefore,  only $\alpha=1/2$ allows divergent $R''(1)$. 
Only in this case does the nontrivial critical behavior 
differ from the prediction of dimensional reduction. 
Since the initial value $R(z)$ of the renormalization group
equation (\ref{1loop}) is an analytic function, the flow of $R''(1)$ should 
diverge for the breakdown of dimensional reduction.

The same discussion for $z=-1$ can be done. The only possible singularity is 
$$
R'(z) = R'(-1) + C(1+z)^{1/2}+ \cdots,
$$
with $C=-R'(1)$.

Next, we consider the renormalization group flow 
of the singular function $R'(z)$.
If we assume an initial coupling 
\begin{equation}
R'(z) =C (1-z)^{\alpha},
\label{singular}
\end{equation}
with $ \alpha > 0 $, 
the renormalization group transformation generates a term 
$
-C^2(1-z)^{2\alpha-1}.
$
If $2 \alpha-1 < \alpha$--namely, 
$\alpha < 1$--the successive transformations produce less power. 
Eventually, the flow generates
a term 
$$
R'(z) \sim (1-z)^{\alpha'},
$$
with $\alpha' < 0$, unless $\alpha=\frac{1}{2}$ or $\alpha \geq 1$.
To avoid the flow to such
unphysical regions, we require a condition
\begin{equation}
\alpha= \frac{1}{2} \ \ \ {\rm or}  \ \  \alpha \geq 1
\label{condition}
\end{equation} 
on the initial function (\ref{singular}). Therefore, the allowed function 
has the same singularity as that of the fixed-point function.

\section{Large-$N$ limit}
\label{Large N limit}

Here, we take the large-$N$ limit in the one-loop renormalization group
with $NR(z)$ finite and redefine $R(z)$ by $NR(z)\rightarrow R(z)$. 
The one-loop $\beta$ function for $R(z)$ becomes 
\begin{eqnarray}
\partial_tR(z)
&=&-\epsilon R(z)+2R'(1)R(z)-zR'(1)R'(z)\nonumber\\
&&+\frac{1}{2}R'(z)^2.
\end{eqnarray}

\subsection{Fixed points}
\label{Fixed points}

Following the method given by Balents and Fisher, \cite{BF} 
we consider the flow equation for $R'(z)$ instead of that for $R(z)$. 
Differentiating the one-loop $\beta$ function with respect to $z$, we have 
\begin{eqnarray}
\partial_tR'(z)
&=&-\epsilon R'(z)+R'(1)R'(z)-zR'(1)R''(z)\nonumber\\
&&+R'(z)R''(z).
\end{eqnarray}
We redefine the parameters
\begin{eqnarray}
R'(z)\equiv \epsilon u(z),\quad
t'\equiv\epsilon t,\quad
u(1)\equiv a,
\end{eqnarray}
and we consider the fixed-point equation 
\begin{eqnarray}
0=(a-1)u(z)-zau'(z)+u(z)u'(z).
\label{FPE2}
\end{eqnarray}
Substituting $z=1$ into Eq. (\ref{FPE2}), we have two cases 
\begin{eqnarray}
a=0,1,
\end{eqnarray}
for finite $u'(1)$.
Solving the differential equation (\ref{FPE2}) for $a=1$, 
we have two nontrivial solutions: 
\begin{eqnarray}
R'(z)=\epsilon
\label{sol11}
\end{eqnarray}
and
\begin{eqnarray}
R'(z)=\epsilon z.
\label{sol22}
\end{eqnarray}
The first one indicates 
\begin{eqnarray}
&&(\Delta_1,\Delta_2)=(\epsilon,0),\\
&&(R'(1),R''(1))=(\epsilon,0).
\end{eqnarray}
Thus, the solution (\ref{sol11}) is the \lq\lq random field solution.\rq\rq 
The second one indicates
\begin{eqnarray}
&&(\Delta_1,\Delta_2)=\left(0,\frac{\epsilon}{2}\right),\\
&&(R'(1),R''(1))=(\epsilon,\epsilon).
\end{eqnarray}
Thus, the solution (\ref{sol22}) 
is not the \lq\lq random field solution\rq\rq 
but the \lq\lq random anisotropy solution.\rq\rq 

For $a=0$, 
the nontrivial solution is 
\begin{eqnarray}
R'(z)=\epsilon(z-1).
\label{sol33}
\end{eqnarray}
This indicates
\begin{eqnarray}
&&(\Delta_1,\Delta_2)=\left(-\epsilon,\frac{\epsilon}{2}\right),\\
&&(R'(1),R''(1))=(0,\epsilon).
\end{eqnarray}
Thus, the solution (\ref{sol33}) is unphysical. This fact indicates that 
the trivial fixed point $R'(z)=0$ is the unique physical solution 
when $a=0$.

\begin{figure}[ht]
\begin{center}
\setlength{\unitlength}{1mm}
\begin{picture}(60, 35)(0,0)
     \put(0,-5){ 
\includegraphics[width=60mm]{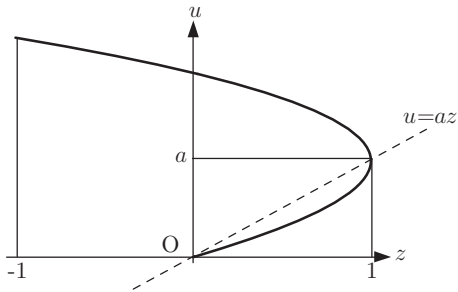}
		}
\end{picture}
\end{center}
\caption{A schematic graph of $u(z)$. 
Since the derivative of $u$ is ill defined on $u=az$, 
the solution terminates on this line. 
The above graph represents two solutions meeting at $(1, a)$. }
\label{graph_u}
\end{figure}

If $a\neq 0,1$, 
\begin{eqnarray}
\frac{du(z)}{dz}=\frac{(a-1)u(z)}{za-u(z)}.
\label{z/u 1DE}
\end{eqnarray}
Taking the inversion, we regard $z$ as a function of $u$. One gets 
\begin{eqnarray}
\frac{dz(u)}{du}=\frac{a}{a-1}\frac{z(u)}{u}-\frac{1}{a-1},
\end{eqnarray}
which is easily integrated. 
Then, we have 
\begin{eqnarray}
z=u-(a-1)\biggl|\frac{u}{a}\biggr|^{a/(a-1)}.\label{z u relation}
\label{NA}
\end{eqnarray}
Because $z(u)$ takes the maximum value 1 at $u=a$, 
$u(z)$ is double valued as we show in Fig. \ref{graph_u}. 
It is seen from Eq. (\ref{z/u 1DE}) that $du/dz$ is ill defined on 
$u=a z$. Therefore the lower branch terminates at the origin, 
so that it should be continued to the region $ -1\leq z < 0$. 
This is possible only if $a/(a-1)=n$ is a positive integer. 
In this case, we have $a = \frac{n}{n-1} > 1$.

%
%

Expanding $u$ around $a$, we have 
\begin{eqnarray}
z
&=&
u-(a-1)\biggl|\frac{u}{a}\biggr|^{a/(a-1)}\nonumber\\
&=&
a+(u-a)-(a-1)\biggl(1+\frac{u-a}{a}\biggr)^{a/(a-1)}\nonumber\\
&\simeq&
1-\frac{1}{2a(a-1)}(u-a)^2.
\end{eqnarray}
Since $-1\le z\le 1$, we have 
\begin{eqnarray}
1-z\simeq\frac{(u-a)^2}{2a(a-1)}\ge 0.\label{cusp1}
\end{eqnarray}
Thus, we find that the fixed point $a$ must satisfy $a(a-1) > 0$.
Since  $a < 0$ gives a negative exponent $ \epsilon + \bar{\eta}  <  0$
of the disconnected
correlation function, $a < 0$ is excluded and we have $a > 1$.
The Schwartz-Soffer inequality  \cite{SS}
$2\eta\ge{\bar{\eta}}$
also requires the other constraint
$a \leq 1$.
Therefore, these nonanalytic solutions (\ref{NA})
are unphysical.

\subsection{Stabilities of the fixed-point solutions}

Next, we study the stability of the fixed points. 
Let $u(z)$ be a fixed-point solution: 
\begin{eqnarray}
0=u(z)(a-1)+u'(z)(u(z)-az).
\end{eqnarray}
We consider an infinitesimal deformation $u(z)\rightarrow u(z)+v(z)$ 
and $a\rightarrow a+b$, with a finiteness 
condition $\sup_{-1 \leq z \leq 1}|v(z)| < \infty$
and $b< \infty$.
We study the behavior of the first order in $v(z)$ and $b$: 
\begin{eqnarray}
&&v(z)(a-1)+u(z)b+v'(z)[u(z)-az]\nonumber\\
&&+u'(z)[v(z)-bz]=\lambda v(z).
\label{v1}
\end{eqnarray}
$\epsilon \lambda$ denotes the eigenvalue of the scaling operator. 
The negative eigenvalue $\epsilon \lambda<0$ indicates that 
the fixed-point solution is stable, 
and the positive eigenvalue $\epsilon \lambda>0$ indicates that 
the fixed-point solution is unstable. 
Normalizing $v(z)$ appropriately, we can take $v(1)=0$ or $v(1)=1$.

\subsubsection{$R'(z)=\epsilon$}
For $a=1$ and $u(z)=1$, Eq. (\ref{v1}) becomes 
\begin{eqnarray}
b+v'(z)(1-z)=\lambda v(z), 
\end{eqnarray}
where $b$ represents $v(1)$ taking $0$ or $1$. 
When $b=0$, 
the solution is 
\begin{equation}
v(z)=C(1-z)^{-\lambda}, 
\end{equation}
where $\lambda<0$ because of the initial condition $b=v(1)=0$. 
On the other hand, when $b=1$, 
a general solution is 
\begin{equation}
v(z)=\left\{
\begin{array}{ll}
\lambda^{-1}+c(1-z)^{-\lambda} \ \ \ \ &(\lambda\neq0),\\
\ln|1-z| \ \ \ \ &(\lambda = 0). 
\end{array}
\right.
\end{equation}
Here the condition $b =1$ requires that $\lambda=1$ and $c=0$. 
Since the singularity of the physical deformation $v(z)$ satisfies the condition 
(\ref{singular}), the allowed value of $\lambda$ is 
\begin{equation}
\lambda=-\frac{1}{2}, \ \ \ 
\lambda \leq -1 \ \ \ {\rm or} \ \ \ \lambda =1. 
\end{equation}
This shows that the fixed-point solution  is singly unstable. 
Note that the cuspy deformation from this fixed point is irrelevant.

\subsubsection{$R'(z)=\epsilon z$}
For $a=1$ and $u(z)=z$, Eq. (\ref{v1}) becomes 
\begin{eqnarray}
v(z)=\lambda v(z). 
\end{eqnarray}
Then, $\lambda=1$, and the fixed point is fully unstable. 
Note that this instability
occurs for large $N$.

\subsubsection{$R(z)=0$}
Since $a=0$ and $u=0$ in this case, 
Eq. (\ref{v1}) is $-v(z)=\lambda v(z)$, 
which means $\lambda = -1$ for any $v(z)$; 
thus the trivial fixed point is fully stable.

\section{Two-loop renormalization group analysis}
\label{Subleading corrections}
Here, we discuss the stability of the fixed point
by solving the eigenvalue problem of the scaling operator 
in the two-loop renormalization group. 

\subsection{The two-loop $\beta$ function}
Recently,
Le Doussal and Wiese calculated 
a two-loop $\beta$ function at zero temperature. 
Independently, Tarjus and Tissier obtained the following consistent result 
\begin{eqnarray}
&&\beta_2[R]\nonumber\\
&&=\frac{1}{2}(N-2)\{(1-z^2)^2R''(z)^3\nonumber\\
&&-(1-z^2)[3z R'(z)-(2+z^2)R'(1)]R''(z)^2 \nonumber\\
&&-2(1-z^2)[R'(z)-z R'(1)]R'(z)R''(z)\nonumber\\
&&+(1-z^2)R'(1)R'(z)^2+4R'(1)^2R(z)\}\nonumber\\
&&-\frac{1}{2}(1-z^2)[(1-z^2)R'''(z)-3z R''(z)-R'(z)]^2\nonumber\\
&&\times[-(1-z^2)R''(z)+z R'(z)-R'(1)]\nonumber\\
&&-\frac{c^2}{2}[(N+2)(1-z^2)R''(z)-(3N-2)zR'(z)\nonumber\\
&&+8K(N-2)R(z)],
\label{2loopbeta}
\end{eqnarray}
where $c= \lim_{z \nearrow 1} \sqrt{1-z^2}R''(z)$ and 
$K=2 \gamma_a$ is an unknown real number. 
As discussed in the one-loop $\beta$ function in Sec. \ref{singularity}, 
the possible singularity 
in the fixed-point function $R'(z) \sim (1-z)^\alpha$ 
and the possible deformation from the fixed point 
are given by $\alpha =\frac{1}{2}$ or
$\alpha \geq 1 $.
Otherwise, the function cannot be a fixed point. 
This condition is preserved exactly also in the two-loop $\beta$ function. 
In the following, we present some corrected results based on the two-loop renormalization group.

\subsection{Fixed point in the two-loop $\beta$ function}
First, we discuss a condition on $N$ and $d$ for the existence of the
fixed point corresponding to the dimensional reduction. 
The two-loop $\beta$ functions for the two coupling constants are
\begin{eqnarray}
\partial_t R'(1)&=&-\epsilon R'(1) +(N-2)R'(1)^2\nonumber\\
&&+(N-2)R'(1)^3,\nonumber\\
\partial_t R''(1)&=&-\epsilon R''(1) +(N+7) R''(1)^2+6R''(1)R'(1)\nonumber \\
&&+R'(1)^2+2(5N+17) R''(1)^3\nonumber\\
&&+6(N+7)R''(1)^2R'(1)\nonumber\\
&&-6(N-5)R''(1)R'(1)^2-(N-4)R'(1)^3.\nonumber\\
\end{eqnarray} 
As in the one-loop case, the flow of 
these coupling constants does not depend on higher derivative couplings.  
The flow is qualitatively the same as the one-loop case depicted in Fig. \ref{flow2}.
The fixed point corresponding to the dimensional reduction has the following correction 
\begin{equation}
R'(1) = \frac{\epsilon}{N-2}-\left(\frac{\epsilon}{N-2}\right)^2.
\end{equation}
The fixed-point equation for $R''(1)$ becomes a cubic algebraic equation.
For the nonexistence of the solution $R''(1)$ in the perturbative region,
this cubic equation has only one real solution.
This condition requires
the negative discriminant 
\begin{equation}
D=-\frac{4}{a^3}\left(c-\frac{b^2}{3a} \right)^3-\frac{27}{a^2}\left( \frac{2b^3}{27 a^2}-
\frac{b c}{3a} +d \right)^2 < 0
\label{discr}
\end{equation}
for the cubic equation $aR''(1)^3+b R''(1)^2+cR''(1)+d=0$.
Expanding the left-hand side of the inequality (\ref{discr}) in $N-18$ and $\epsilon =d-4$, we obtain  
\begin{equation}
N < 18- \frac{49}{5} \epsilon.
\end{equation}
This condition determines the region where the dimensional reduction breaks. 
The condition on $N$ for the nonexistence of 
the solution $R''(1)$ is corrected from
$N < 18$ in the two-loop order. 
This condition is identical to the existence condition of a suitable 
cuspy fixed point obtained by Le Doussal and Wiese, \cite{DW} which is
consistent also with the phase diagram obtained by Tarjus and Tissier.
In higher dimensions, dimensional reduction occurring is more likely.

\subsection{Fixed point in the double expansion}
The two-loop $\beta$ function (\ref{2loopbeta}) 
enables us to calculate the higher-order 
corrections to the fixed point corresponding to dimensional reduction. 
For $N \geq 18-\frac{49}{5} \epsilon$,
this fixed point may exist, and it can be obtained 
in a double expansion with respect to $1/N$ and $\epsilon$
to several orders. We expand the fixed point up to necessary orders 
to discuss its stability.  
\begin{eqnarray}
R(z)&=&\frac{\epsilon}{N}\left(z-\frac{1}{2}\right)
+\frac{\epsilon}{N^2} \left(\frac{1}{2}z^2 + z \right)\nonumber\\
&&+\frac{\epsilon^2}{N^2} \left(-\frac{1}{2}z^2 -\frac{1}{2} \right)+\cdots. 
\end{eqnarray}
We study the stability of this fixed point in the scaling operator.
In the two-loop analysis, we define the deviation function from this fixed point by 
$$
\delta R'(z) = \frac{\epsilon}{N} v(z).
$$
We can calculate the correction of these higher orders to the eigenvalue 
equation for the deviation: 
\begin{eqnarray}
&&(1-\epsilon)(1-z)^2(1+z)v''(z)\nonumber\\
&&+\left[N-4z-2+\epsilon(2+4z) \right](1-z)v'(z)\nonumber \\
&&+\left[ 2(1-\epsilon)z-N\lambda\right] v(z)+\left[N-2+\epsilon(4-z)\right]v(1)=0.\nonumber\\ 
\label{eigenvalueeq2loop}
\end{eqnarray}
As discussed in the large-$N$ limit, 
we can determine the eigenvalues $\epsilon \lambda$. 
First, we study the equation for $v(1)=0$.
The solutions of this equation have regular singular points $z=1$ and 
$-1$ for the interval $-1 \leq z \leq 1$.  
Therefore, we can  obtain the solutions in the following expansion forms: 
\begin{equation}
v(z) = (1-z)^{\alpha} \sum_{n=0} ^\infty a_n (1-z)^n
\label{alpha}
\end{equation}
around $z=1$ and
\begin{equation}
v(z) = (1+z)^{\beta} \sum_{n=0} ^\infty b_n (1+z)^n
\label{beta}
\end{equation}
around $z=-1$. Substituting these forms into the eigenvalue equation (\ref{eigenvalueeq2loop}), we require that the coefficients of the lowest order vanish. 
This requirement gives the indicial equations for the 
exponents $\alpha$ and $\beta$
\begin{eqnarray}
2\left(1-\epsilon \right)\alpha^2
-\left(N-4+4\epsilon \right)\alpha
+2-2\epsilon-N \lambda=0,
\end{eqnarray}
\begin{eqnarray}
2(1-\epsilon)\beta^2+N \beta=0,
\end{eqnarray}
which have the solutions 
\begin{eqnarray}
\alpha&=&\alpha_{\pm}(\lambda)\nonumber\\
&\equiv&\frac{N-4(1-\epsilon) 
\pm \sqrt{N^2-8N(1-\epsilon)(1- \lambda)}}{4(1-\epsilon)},  \label{solalpha} \\ 
\beta&=&-\frac{N}{2(1-\epsilon)} \ \ {\rm or} \ \  0. \label{solbeta}
\end{eqnarray}
The coefficient of an arbitrary order satisfies 
the following recursion relation: 
\begin{eqnarray}
&&k[2(1-\epsilon)(k+2\alpha)-N+4(1-\epsilon)] a_k \nonumber\\
&&-(1-\epsilon)(\alpha+k)(\alpha+k+1)a_{k-1} =0 ,\nonumber
\end{eqnarray}
for $k=1, 2, 3, \ldots.$
The solution of this recursion relation is 
\begin{equation}
a_k= \frac{(1+\alpha)_k (2+\alpha)_k}{(3+2 \alpha -N/(2-2\epsilon))_k} 
\frac{a_0}{2^k k!},
\end{equation}
where the symbol $(\cdots)_k$ denotes
\begin{equation}
(x)_k = \Gamma(x+k)/\Gamma(x).
\end{equation}
The solutions can be expressed in terms of the Gaussian hypergeometric function
\begin{eqnarray}
&&v(z, \alpha)\nonumber\\
&&=(1-z)^{\alpha}\nonumber\\
&&\times _2F_1 \left(1+\alpha, \ 2+\alpha; 
\ 3+2\alpha-\frac{N}{2(1-\epsilon)} ; 
\ \frac{1-z}{2} \right),\nonumber 
\end{eqnarray}
where the generalized hypergeometric function 
is defined by the following series expansion
\begin{eqnarray}
&& _mF_n \left(x_1,x_2,\ldots,x_m; y_1, y_2, \ldots, y_n; z \right)\nonumber\\
&&\equiv\sum_{k=0} ^\infty \frac{(x_1)_k(x_2)_k \cdots (x_m)_k}{(y_1)_k(y_2)_k 
\cdots (y_n)_k }  
\frac{z^k}{k!} .
\end{eqnarray}
The general solution is given by the linear combination of the two independent solutions
\begin{equation}
v(z) = a_0^+ v(z, \alpha_+(\lambda))+ a_0^- v(z, \alpha_-(\lambda) ),
\end{equation}
where $\alpha_\pm(\lambda)$ is defined by Eq. (\ref{solalpha}).
In general, this solution has a singularity at $z=-1$ corresponding to
$\beta=-\frac{N}{2(1-\epsilon)}$ given in Eq. (\ref{solbeta}).
To obtain physical solutions, we eliminate the singularity at $z=-1$ by the suitable choice of
$a_0^{\pm}$ for a requirement $|v(z)| < \infty$. 
Also the finiteness of the flow requires 
$\alpha_-(\lambda) = \frac{1}{2}$ or $\alpha_{-}(\lambda) \geq 1$ 
by the condition (\ref{condition}); 
then, we obtain the possible eigenvalues $\epsilon \lambda$ given by
\begin{eqnarray}
&&\lambda = -\frac{1}{2} + \frac{9-9\epsilon}{2N} + \cdots, \\
&&\lambda \leq -1 + \frac{8-8 \epsilon}{N} +\cdots.
\end{eqnarray}
Next we consider the case $v(1) \neq 0$.
The existence of the solution given in the expansion requires vanishing the 
coefficient of $v(1)$, which determines $\lambda$: 
\begin{equation}
\lambda = \lambda_1=1 +\frac{\epsilon}{N} + \cdots.
\label{lambda}
\end{equation}
The corresponding special solution is represented in terms 
of the generalized hypergeometric function 
\begin{eqnarray}
&&v_1(z)\nonumber\\
&&=\frac{\epsilon }{2-2\epsilon}+\frac{2-3\epsilon}{2-2\epsilon}\nonumber\\ 
&&\times _3F_2 \left(1,1,2,;1-\alpha_+(\lambda_1),1- \alpha_-(\lambda_1); 
\frac{1-z}{2}\right).
\label{relevant}
\end{eqnarray}
The finite solution is given by a linear combination of this 
special solution $v_1(z)$
and a solution of the homogeneous
equation with $v(1)=0$ and with the $\lambda$ given in Eq. (\ref{lambda}): 
\begin{equation}
v(z) = a_0 ^+ v(z, \alpha_+(\lambda_1) ) +  v_1(z).
\label{uniquerelevant}
\end{equation}
Since the coefficient $a_0 ^+$ should be chosen 
for the cancellation of the singularity at $z=-1$,
this relevant mode is unique.
Therefore, the fixed point is singly unstable; 
then, dimensional reduction can be observed 
in the critical exponents $\eta$ and $\bar{\eta}$ for
 $N \geq 18 -\frac{49}{5} \epsilon$. 
This result agrees with
a nonperturbative renormalization group obtained 
by Tarjus and Tissier \cite{TT} and a simple $1/N$ expansion. \cite{SMI} 
All eigenfunctions are singular at $z=1$.
It is interesting that the eigenfunction (\ref{uniquerelevant}) has
an essential singularity at $1/N=0$. This is because of the fact that 
we solve the eigenvalue equation 
nonperturbatively without a $1/N$ expansion after the derivation of the eigenvalue equation.
Note that the two-loop renormalization group
is useful to show the existence of the relevant mode.
The limit $\epsilon \rightarrow 0$ 
of the relevant mode (\ref{relevant}) corresponds to the eigenfunction
of the one-loop scaling operator
with the subleading correction in the $1/N$ expansion. 
The series expansion for the relevant mode seems 
to be ill defined, since 
$$
\lim_{\epsilon \rightarrow 0} [1-\alpha_+(\lambda_1)]=2-\frac{N}{2}
$$
is a negative integer for even $N \geq 6$.
This apparent ill definition disappears by 
the two-loop or other higher-order corrections.

Here, we comment on the infinitely many relevant modes 
pointed out by Fisher. \cite{Fi} 
To compare our result to Fisher's one-loop
renormalization group analysis, we take the limit $\epsilon \rightarrow 0$ 
in the solution  (\ref{solalpha}).
They are included in the following series  
$$
\alpha_+(\lambda_k)= k-1,  \ (k=3, 4, 5, \ldots ) \ \ \ {\rm and} \ \ \ a_0^-=0.$$ 
These belong to the eigenvalues $\epsilon \lambda_k$ given by
$$
\lambda_k =  1-k+ \frac{2k^2}{N}+ {\rm O}\left(\frac{1}{N^2}\right),
$$  
which are positive for sufficiently large $k$. 
These agree with the eigenvalues (\ref{eigen k}) obtained by Fisher. Since 
these relevant modes diverge at $z=-1$, 
we have eliminated them as unphysical modes, as discussed above.

\section{Higher-order calculation}
\label{higher}

Here, we calculate higher-order correction in the $1/N$ expansion
to the results. 
The fixed point corresponding to the dimensional reduction 
can be obtained in a double expansion 
with respect to $1/N$ and $\epsilon$
to several orders: 
\begin{eqnarray}
R(z)&=&\frac{\epsilon}{N}\left(z-\frac{1}{2}\right)
+\frac{\epsilon}{N^2} \left(\frac{1}{2}z^2 + z \right)\nonumber\\
&&+\frac{\epsilon^2}{N^2} \left(-\frac{1}{2}z^2 -\frac{1}{2} \right) \nonumber \\
&&+\frac{\epsilon}{N^3} \left(z^3+\frac{5}{2}z^2-4z+\frac{9}{2}\right)\nonumber\\
&&+\frac{\epsilon^2}{N^3} \left(-\frac{7}{4}z^3-\frac{15}{4}z^2+
\frac{35}{4}z-\frac{33}{4}\right) + \cdots. 
\end{eqnarray} 
We study the stability of this fixed point in the scaling operator.
We can calculate the correction of these higher orders to the eigenvalue 
equation for the deviation: 
\begin{eqnarray}
&&-\frac{\epsilon}{N}(1-z)^3 (1+z)^2v'''(z)  \\
&&+\left[1-\epsilon +\frac{1}{N}(3+2 z)+\frac{3 \epsilon}{2N} (1+3 z)\right]\nonumber\\
&&\times(1-z)^2(1+z)v''(z)\nonumber\\
&&+\left[N-4z-2+\epsilon(2+4z)-\frac{1}{N}(7 + 17 z + 10 z^2)\right.\nonumber\\
&&\left.+\frac{\epsilon}{4N}(31 + 27 z + 18 z^2) \right]
(1-z)v'(z) \nonumber \\
&&+\left[2(1-\epsilon)z+\frac{1}{N}(6z^2+8z)\right.\nonumber\\
&&\left.-\frac{\epsilon}{2N}(15z^2+21z+2)-N\lambda\right]v(z)\nonumber\\
&&+\left[N-2+\epsilon(4-z)-\frac{1}{N}(7+4z+3z^2)\right.\nonumber\\
&&\left.+\frac{\epsilon}{N}\biggl(\frac{59}{4}+7z-\frac{3z^2}{4}\biggr)\right]v(1)=0. \nonumber
\end{eqnarray}
As discussed in the previous section, we can determine the eigenvalues $\epsilon \lambda$. 
First we discuss the solution with $v(1)=0$.
To obtain finite solutions in the expansion (\ref{alpha}) 
around $z=1$ and (\ref{beta}) around $z=-1$,
we should require that the coefficients of the lowest order vanish. 
These conditions give the indicial equations for the 
exponents $\alpha$ and $\beta$ with two-loop corrections: 
\begin{eqnarray}
&&\frac{4 \epsilon}{N} \alpha^3
+2\left(1-\epsilon +\frac{5}{N}\right)\alpha^2\nonumber\\
&&-\left(N-4+4\epsilon-\frac{24}{N} 
+\frac{23\epsilon}{N}\right)\alpha\nonumber\\
&&+2-2\epsilon+\frac{14}{N}-\frac{19 \epsilon}{N}-N \lambda=0, \nonumber \\
&&-\frac{8\epsilon}{N} \beta^3+
4\left(1-\epsilon+\frac{1}{N}+\frac{3\epsilon}{N}\right)\beta^2\nonumber\\
&&+\left(2N -\frac{4}{N}+\frac{7\epsilon}{N}\right)\beta=0.
\end{eqnarray}
The second equation has three solutions for $\beta=\beta_+ > 0$,  $\beta=0$, 
and $\beta = \beta_- <0$. Since the solution with $\beta_-$ diverges at $z=-1$,
we should eliminate this solution.
To this end, we need two independent solutions which are finite at $z=1$.
The existence of two solutions of $\alpha= \frac{1}{2}$ or 
$\alpha \geq 1$ requires the eigenvalue $\epsilon \lambda$ given by
\begin{eqnarray}
&&\lambda = -\frac{1}{2} + \frac{9-9\epsilon}{2N} +\frac{57-60 \epsilon}{2N^2} + \cdots, \\
&&\lambda \leq -1 + \frac{8-8 \epsilon}{N}+\frac{48-38\epsilon}{N^2} +\cdots.
\end{eqnarray}
Next we consider the case $v(1) \neq 0$.
The existence of the solution given in the expansion requires vanishing the 
coefficient of $v(1)$, which determines
\begin{equation}
\lambda = 1 +\frac{\epsilon}{N} + \frac{2 \epsilon}{N^2} + \cdots.
\label{lambda2}
\end{equation}
The finite solution is given by a linear combination of this special solution 
of the inhomogeneous equation and a solution of the homogeneous
equation with $v(1)=0$ and with the $\lambda$ given in Eq. (\ref{lambda2}).
Therefore, the fixed point is singly unstable; 
then, dimensional reduction can be observed 
in the critical exponents $\eta$ and $\bar{\eta}$ for $N \geq 18 -\frac{49}{5} \epsilon.$

\section{Discussions}
\label{Conclusion}

In this paper, we have studied the stability of the 
fixed points in the renormalization group for the
 random field O($N$) spin model in $4+\epsilon$ dimensions
for sufficiently large $N$. We argue physical conditions on the
random anisotropy function. 
This argument enables us to solve the eigenvalue problem 
for the infinitesimal deviation from all fixed points. 
In the large-$N$ limit, all nontrivial 
fixed points are unstable or unphysical, except 
the fixed point corresponding to dimensional reduction.

We also discuss the stability of this fixed point 
on the basis of the two-loop renormalization group 
obtained by Le Doussal and Wiese \cite{DW} and 
Tissier and Tarjus. \cite{TT2} These corrected results are
essentially the same as those in the large-$N$ limit.
The fixed point corresponding to 
dimensional reduction cannot exist for $N < 18 -\frac{49}{5} \epsilon.$
This condition on $(d, N)$   
agrees with the phase diagram obtained by Tarjus and Tissier \cite{TT}
in a nonperturbative renormalization group and also with an
existence condition of the cuspy fixed point
obtained by the Le Doussal and Wiese. \cite{DW} 

We derive the eigenvalue equation of the scaling operator in the 
double expansion in 
$1/N$ and $\epsilon$ on the basis of the two-loop renormalization group, and
solve it nonperturbatively. 
We show that the unique singly unstable fixed point 
gives the critical exponents $\eta$ and $\bar{\eta}$ 
predicted by dimensional reduction. 
This result agrees with that obtained by Tarjus and Tissier in the 
nonperturbative renormalization group \cite{TT} and also with 
the stability of 
the replica-symmetric saddle-point solution in the $1/N$ expansion. \cite{SMI} 
Here, we summarize the physical insights obtained from our mathematical results
for the random field O($N$) spin model
in $4+\epsilon$ dimensions at zero temperature.
There are two phases for sufficiently weak randomness. The fully stable
fixed point $R(z)=0$ makes the extended ferromagnetic phase.
This phase is specified by the nonzero constant order parameter.  
The correlation of the spin fluctuation
obeys the power law decay 
with the same exponent as that in the trivial
Gaussian model.
In the disordered phase,
the order parameter vanishes 
and the spin correlation becomes short ranged.
For $N \geq 18 -\frac{49}{5} \epsilon$,
the phase transition between these two phases 
is governed by the singly unstable fixed point
corresponding to dimensional reduction. Therefore, at any phase 
transition point on the phase boundary, the 
critical exponents predicted by dimensional reduction
should be observed universally. On the other hand, for 
$N < 18 -\frac{49}{5} \epsilon,$ the fixed point corresponding to 
dimensional reduction disappears. In this case, it is believed that
a cuspy nonanalytic fixed point $R'(z)\sim(1-z)^{1/2}$ 
governs the phase transition between the 
ferromagnetic and disordered phases. 
The breakdown of dimensional reduction 
is observed in the critical exponents. \cite{Fe}

It is important to explore a breakdown of dimensional reduction
in some other observables even for $N \geq 18 -\frac{49}{5} \epsilon$.
For this problem, 
the statement on the fixed point 
by Tarjus and Tissier is interesting. \cite{TT} 
They argue that the singly unstable fixed point has a weak singularity
\begin{equation}
R'(z) \sim (1-z)^{N/2},
\label{weaksingularity}
\end{equation}
at $z=1$ for large $N$. 
Even if the fixed point corresponding to dimensional reduction 
has this singularity, 
we cannot find such an essential singularity in the $1/N$ expansion. 
Therefore, it is possible that the obtained fixed point in the $1/N$ expansion 
is not analytic and has the weak singularity (\ref{weaksingularity}).  
\begin{acknowledgments}
This work is supported by a Grant-in-Aid for Scientific Research Program (No. 18740242) 
from the Ministry of Education, Science, Sports, Culture and Technology of Japan. 
\end{acknowledgments}

\end{document}